\documentclass[sigconf]{acmart} 

\usepackage[indentfirst=false,font={itshape},begintext=``,endtext='']{quoting}
\usepackage{subcaption}
\usepackage{hyperref}
\usepackage[most]{tcolorbox}
\newtcolorbox{summarybox}{
    enhanced,
    breakable,
    sharp corners,
    boxrule=0.3pt,
    colback=white,
    colframe=black!40,
    coltitle=black,
    fonttitle=\bfseries,
    toptitle=2mm,
    bottomtitle=2mm,
    colbacktitle=white,
}

\AtBeginDocument{%
  \providecommand\BibTeX{{%
    \normalfont B\kern-0.5em{\scshape i\kern-0.25em b}\kern-0.8em\TeX}}}

\acmConference[GAS '22]{6th International ICSE Workshop on Games and Software Engineering:  Engineering fun, inspiration, and motivation}{June 03--05, 2018}{Pittsburgh, PA, USA}

\makeatletter

\usepackage{amssymb}
\newcommand{\ygg@basicalert}[2]{\textcolor{red}{\fbox{\bfseries\sffamily\scriptsize#1}{\sf\small$\blacktriangleright$\textit{#2}$\blacktriangleleft$}}}
\newcommand{\YANN}[1]{\ygg@basicalert{YANN}{#1}}

\copyrightyear{2022}
\acmYear{2022}
\setcopyright{acmcopyright}\acmConference[GAS'22]{6th International ICSE Workshop on Games and Software Engineering: Engineering fun, inspiration, and motivation}{May 20, 2022}{Pittsburgh, PA, USA}
\acmBooktitle{6th International ICSE Workshop on Games and Software Engineering: Engineering fun, inspiration, and motivation (GAS'22), May 20, 2022, Pittsburgh, PA, USA}
\acmPrice{15.00}
\acmDOI{10.1145/3524494.3527623}
\acmISBN{978-1-4503-9293-8/22/05}

\begin{document}

\title{What Makes a Game High-rated?\\Towards Factors of Video Game Success}
\author{Gabriel C. Ullmann}
\email{g_cavalh@live.concordia.ca}
\orcid{0000-0002-3274-0789}
\affiliation{
  \institution{Concordia University}
  \city{Montreal}
  \state{Quebec}
  \country{Canada}
}

\author{Cristiano Politowski}
\email{c_polito@encs.concordia.ca}
\orcid{0000-0002-0206-1056}
\affiliation{
  \institution{Concordia University}
  \city{Montreal}
  \state{Quebec}
  \country{Canada}
}

\author{Yann-Ga\"el Gu\'{e}h\'{e}neuc}
\email{yann-gael.gueheneuc@concordia.ca}
\orcid{0000-0002-4361-2563}
\affiliation{
  \institution{Concordia University}
  \city{Montreal}
  \state{Quebec}
  \country{Canada}
}

\author{Fabio Petrillo}
\email{fabio@petrillo.com}
\orcid{0000-0002-8355-1494}
\affiliation{
  \institution{Université du Québec à Chicoutimi}
  \city{Chicoutimi}
  \state{Quebec}
  \country{Canada}
}

\begin{abstract}
As the video game market grows larger, it becomes harder for games to stand out from the crowd. Launching a successful game encompasses different factors, some of which are not well-known. In this paper, we investigate some factors that affect game scores, considering high-rated video games from a dataset of 200 projects. Results show that larger team sizes are often linked to higher scores. On the other hand, the level of freedom given to developers, as well as genre, graphical perspective, game modes and platforms do not correlate to score. Additionally, teams from successful games also experience more crunch time while fewer problems with schedule and budget allocation. Further analysis shows that team, technical, and game design factors should be the main focus of the game developers. 

\end{abstract}


\maketitle




\section{Introduction}

The game industry has generated US\$174 billion in worldwide revenue in 2020. This is a growth of 19.6\% compared to the previous year\footnote{\url{https://newzoo.com/key-numbers}}. While this shows that the video game market is growing, this does not mean every game partakes in its growth.

The game market is large. Platforms such as Steam receive hundreds of new game releases every month\footnote{\url{https://steamspy.com/year/} and \url{https://bit.ly/2PTkm1H}}. Standing out from this crowd and making a profit, for a game, is a hard endeavor. A recent example of the difficulties in developing a video game is the case of ``No Man's Sky''. Crowdfunded and produced by the indie studio ``Hello Games'', it suffered from strong criticism for not delivering promised features \cite{chien20}, translating into users massively asking for a refund\footnote{\url{https://www.forbes.com/sites/erikkain/2016/08/29/gamers-have-every-right-to-push-for-no-mans-sky-refunds}}. Similar situations also happen with AAA studios, as in the cases of ``Aliens: Colonial Marines'' \cite{keogh18} and ``Cyberpunk 2077'' \footnote{\url{https://nyti.ms/2R2cGL3}}, which were expected to generate large profits but the final product did not meet the expectations of the players.

Some of the subjective factors leading to the successes or failures of games are well-known, e.g., the game is polished, the gameplay loop is fun, etc. 
But others are less visible or known. Yet, having explicit factors and their relationships with games' successes and failures would help game developers. Consequently, in this paper, we answer the following research question:
\textbf{Which factor better describes high-rated games?}

To answer this question, we collect data from 200 video game projects and analyzed a set of different factors: \textit{team size}, \textit{levels of independence}, \textit{game genre}, \textit{game platform}, \textit{graphical-perspective}, \textit{game mode} and \textit{development problems}. We also collected numeric scores for each game using ``Metacritic'' and other review aggregators as sources. We used this data to investigate which of these factors affect the success of the games. 

We used the numeric scores of the games as a proxy for their successes. They are relevant measures of success from and for both the public and the game industry. For game studios, high scores mean good sales and also potential financial incentives\footnote{For example, in 2012, Chris Avellone, the creative director of the open-world RPG ``Fallout: New Vegas'', declared that the publisher ``Bethesda Softworks'' refused to pay a bonus to the studio ``Obsidian Entertainment'' because the game missed the target Metascore of 85 by 1 point (\url{https://bit.ly/31L0L6t}).}. For the public, it is a way to filter games by their quality.

Our results show that larger teams are often related to higher-rated games. Higher-rated games also present more occurrences of crunch time and fewer problems with scope, budget, and cutting features. However, there is no strong correlation of any of the factors with the game scores. In a further analysis on the high-rated games, in which we compared the occurrence of the most common problems, we confirmed a previous study \cite{politowski20} that reported that \textit{team}, \textit{technical}, and \textit{game-design} factors should be the main focus of the game developers. This answer may help video game project managers and developers to better plan and monitor the projects. 

The paper is structured as follows. Section \ref{sec:method} describes the methods of data collection and inference. Section \ref{sec:results} discusses the results we found by analyzing the dataset and the relations between its variables. Section \ref{sec:discussion} discusses the results. Section \ref{sec:threats} lists the threats to validity. Section \ref{sec:related} shows the related works. Section \ref{sec:conclusion} concludes the paper and discusses future work.

\section{Method}
\label{sec:method}

This work uses seven factors to analyze the projects in the dataset: \textit{team size}, \textit{levels of independence}, game \textit{genre}, game \textit{mode}, game \textit{platform}, graphical \textit{perspective}, and \textit{problem types}. We compare each one of these factors with the game's score, our main metric. We divide the projects according to a binary score classification: high-rated and low-rated games.

We use the dataset of 200 video game projects and their problems, presented by Politowski \textit{et al.} \cite{politowski20}. Besides the problem types, all other factors are new. They were chosen firstly because they were readily available on public sources, such as Metacritic, project postmortems and game store pages. Additionally, the way different game genres, modes and platforms influence players and the market as a whole have been subject of previous studies \cite{joseph14, lindstrom20, tekofsky17}. Therefore, it is relevant to understand how high and low rated games are distributed in each of these contexts.

Postmortems are used by game developers to share information about their projects in the form of ``war stories''. These informal texts summarize the developers' experiences with their games and are often written by managers or senior developers \cite{Callele2005, Washburn2016} right after their games launched. They often include sections about ``What went right'' and ``What went wrong'' during the game development:

\begin{itemize}
\item \emph{``What went right''} discusses Project Management processes and decisions that helped the project.
	
\item \emph{``What went wrong''} discusses difficulties, pitfalls, and mistakes experienced by the development team in the project.
\end{itemize}

\subsection{Scores}
\label{sub:score}

Metacritic classifies games using two metrics: the Metascore, based on posts from popular game review websites, and the User Score, based on user input provided directly via Metacritic \footnote{\url{https://www.metacritic.com/about-metacritic}}. For this work, we chose the Metascore as our main metric because it is based on opinions from critics through channels that are well-known to the general public. User Scores are posted anonymously and are thus vulnerable to ``review bombing''. While Metacritic now enforces a review waiting period to avoid disproportionately low reviews from players, such a problem still cannot be completely avoided \footnote{\url{https://kotaku.com/metacritic-will-now-make-users-wait-36-hours-before-pos-1844421321}}.

The Metascores were manually collected from the site, using its search functionality. However, not all games in our dataset had this score. A minimum of 4 scores from critics are needed for the Metascore to be generated and, in the case of less popular games, this number is often not reached. For such games, we used digital game store scores as a replacement. For games available on PC, we used Steam as the primary source. Apple App Store and Google Play were queried when the game in question had mobile versions. Out of 200 game scores, 181 were obtained directly via Metacritic.

We normalized the scores on a 0--100 scale. For scoring systems that considered 1--5 stars (e.g., Google Play), 5 was considered equivalent to 100, and the relative score was obtained using a simple cross product.

Steam has no numerical score system, so we took Google as a reference, which converts the Steam scoring system to a 1-10 scale. We discovered this relation by observing how results are displayed on Google and what ``label'' is given by Steam to this same score. It goes as follows:

\begin{itemize}
  \item 100 = Overwhelmingly Positive
  \item 90 = Very Positive
  \item 80 = Positive
  \item 70 = Mostly Positive
  \item 60 = Mixed
  \item 50 = Mostly negative
  \item $<$40 = Very/Overwhelmingly Negative (we found no games in this category)
\end{itemize}

\subsection{Team Sizes}
\label{sub:team}

We found the exact number of members in the team in 170 of 200 postmortems. For the remaining 30 projects, we inferred the number based in the postmortem author's description and information we could find in other sources.







For example, if there was no data box in a postmortem\footnote{Summary that appears at the end of some Gamasutra postmortems.}, then we considered the numbers cited in the text. If the developer stated or implied the game was done solo, the team size was considered as 1.

If there was no exact number in the text, then we looked for sources in other media, like articles or press releases. For example, we used the data from the game's credits in the MobyGames repository\footnote{MobyGames is a video game data repository created in 1999 that ``has amassed a database of over 33,000 games, comprising information including developer credits, screenshots, trivia, release dates, platforms and aggregated reviews''(\url{https://bit.ly/3mgOmk6})}.




Finally, if no data was found in the media outlets, then we considered postmortem statements. For example, if a postmortem author described the team size it as ``small'' or ``large'', we have taken this into consideration. However, this information alone was not used as a verdict to decide the team size of the project. For example, for attesting a group as ``small'', we considered evidences such as the project being ``indie'', being sold for a small price, or getting little to no media attention.


\subsection{Levels of Independence}
\label{sub:level-indep}

We adapted the framework defined by Garda and Grabarczyk \cite{Garda}, which delimits \textit{creative}, \textit{financial}, and \textit{publishing} independence. We defined an ordinal scale metric called \textbf{level of independence} to represent how much freedom the development team had. From zero to three (0-3), the bigger the number, the more independent the project is. For example, projects that present creative, financial, and publishing independence have a sum of 3 and therefore are categorized as the ``most independent''. We categorized projects that describe no types of independence (sum equals zero) as ``least independent''. Other cases were categorized as ``partially independent''.

\paragraph{Creative independence}  When a game was self-published or when the postmortem author stated that the creative director or development team as a whole had freedom to propose and execute their own ideas, we considered the game as creatively independent. However, if the game described in the PM was a port, a continuation that was not regarded by its creators as a significant evolution in a franchise, or when the PM author mentioned creative conflicts between developers and publishers, we considered the project did not have creative independence.

\begin{quoting}[endtext='' -- Brutal Legend by Caroline Esmurdoc]
Armed now with a proprietary game engine, a robust tools pipeline, a talented and experienced staff, and the creative freedom and corporate mandate to innovate. 
\end{quoting}

\paragraph{Financial independence} We considered that a game had financial independence if the team used their own resources to fund the development of the game. In the case of a single person or loose group of individuals, these resources may be personal earnings, savings, or credit. In the case of a company, the funding may come from crowdfunding, previous games sales, services provided to other parties, or from a parent company. External funding, coming from angel investors, government grants, contractors, or third-party publishers, does make a game financially independent. Finally, when there was no comments regarding funding, we inferred that the game was self-funded if it was also self-published. Otherwise, we classified the game as financially dependent on its publisher.

\begin{quoting}[endtext='' -- NyxQuest: Kindred Spirits by Rob de Lara]
We decided to partner and fund the game with our own money. In the end this has worked very well, because we managed to do a good looking game in the time we had, and with scarce resources.
\end{quoting}

\paragraph{Publishing independence} If the game was self-published, we considered it as an independent. When the publisher is a parent company of the studio that developed the game, we made the same assumption. When the game was published by a third-party company, we did not consider the game as independent of its publisher.

\begin{quoting}[endtext='' -- Knightly Adventure by Doyon Kim]
As a self-publisher it's key to invest in relationships -- there are ten million other games out there so you can't expect everyone to immediately realize why YOUR game is so special compared to everyone else's. You have to make your case well and respect your partners to get traction when you're an indie developer.
\end{quoting}

\subsection{Game Genres}
\label{sub:genre}

We define a set of game genres using an existing classification \cite{politowski20}. We obtained the genres for each game in our dataset using Wikipedia. If a genre in Wikipedia was not in our set, then they were extrapolated from the definition given by the postmortem authors through examples, texts, or screenshots. If Wikipedia listed more than three genres for a given game, we chose to keep only the first three, to avoid excessive segmentation.

\subsection{Graphical Perspectives}
\label{sub:perspective}

For the graphical perspective analysis, we are not considering the projection rendered by the game engine, but the game-world view as experienced by the player. Thus, we considered 2D games created on a 3D engine as 2D. For 2.5D, we followed a previous definition \cite{xiaoguang06} that ``2.5D is a simplified 3D (x,y,z) surface representation that contains at most one depth value (z-direction) for every point in the (x, y) plane''. Therefore, we considered any 3D game with only one point of view (fixed-angle camera) as 2.5D.

\subsection{Platform, Game Modes, and Problem Types} \label{sub:plat-mod-prob}
\textbf{Platform} types are \textit{PC}, \textit{console}, and \textit{mobile}, the latter encompassing games designed for all kinds of handheld devices \cite{politowski20}.

\subsection{Game Platforms}
For the game \textbf{modes}, we considered \textit{singleplayer}, \textit{multiplayer} (which includes local network or multiple controller play only), and \textit{online} (Internet multiplayer only) \cite{politowski20}.

\subsection{Problem Types}
The \textbf{problems types} consist in a list of 20 items \cite{politowski20}, grouped in three categories (production, management, and business). 

We analyzed the ``What went right'' (WWR) and ``What went wrong'' (WWW) section of the PMs of each of the high-rated games, collecting quotes that best summarised the strongest points described by the authors. 

After gathering these quotes, we related them with the problems types that they described. We assigned multiple problems by postmortem quote based on what was most closely related to the authors' descriptions.

For postmortems that did not have a clear division between ``What went right'' and ``What went wrong'', we considered the entire text but only collected quotes that described factors that described solutions or factors for success, not problems.

\section{Results}
\label{sec:results}

This section shows the general analysis of the dataset. None of the variables seem to explain the game's success. However, there are some other findings described in the following sections.

\subsection{Scores}
\label{sub:res-score}

The score was in $[40, 97]$ with a median of 79, which indicates that higher scores are most common in the dataset. The mean value is 74.88 and the standard deviation is 14.14. The values below 45 are considered outliers. These occurred only for a few games for which no score information could be found.

\begin{figure}[ht]
   \centering
    \includegraphics[width=.6\linewidth]{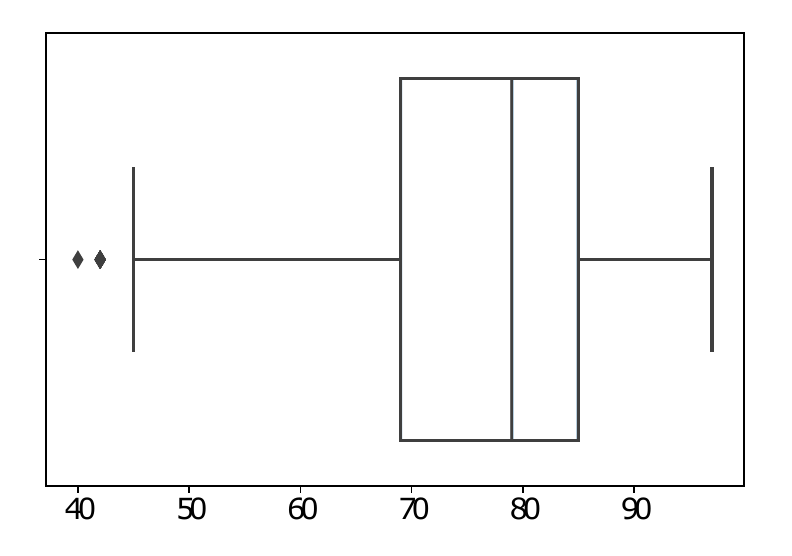}
    \caption{Score distribution}
    \label{fig:boxplot-score}
\end{figure}

According to their scores, the games were divided two groups: \textbf{high-rated} and \textbf{low-rated}. This division was made using the \textit{boxplot quartiles} as reference (\autoref{fig:boxplot-score}). Projects with scores greater than 85 (Q4) were considered high-rated, while the remaining were considered low-rated: out of 200 games, 48 were high-rated and 152 low-rated.

\subsection{Team Sizes}
\label{sub:comp-team}

\begin{figure}[ht]
    \centering
    \includegraphics[width=1\linewidth]{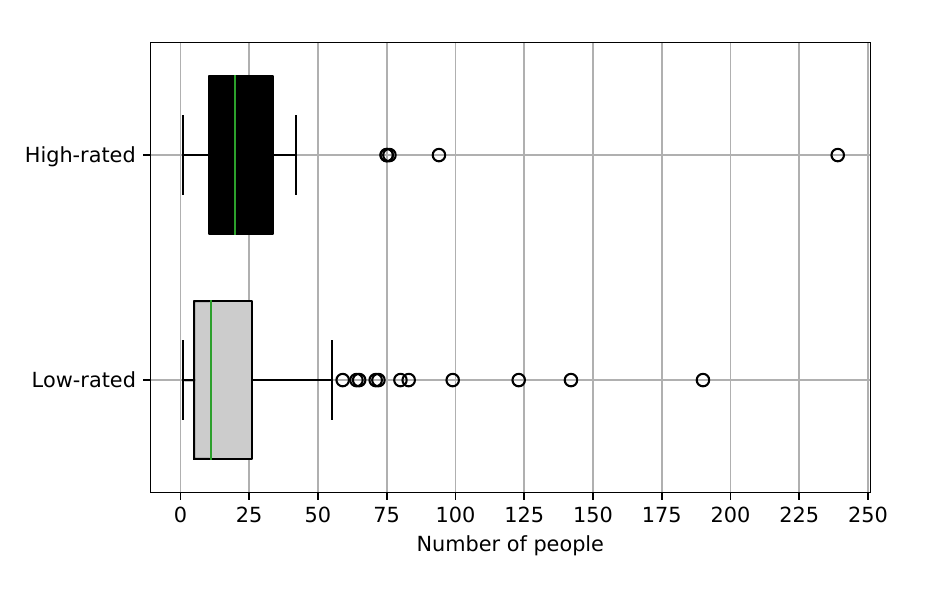}
    \caption{Project's \textbf{team size} of the high-rated games in comparison with the low-rated ones.}
    \label{fig:team}
\end{figure}

\autoref{fig:team} shows the team sizes of the games. For low-rated games, the median team size is 11, while for high-rated ones it is 20. Yet, there are more outliers (above the Q4) for the low-rated games.

\subsection{Levels of Independence}
\label{sub:comp-indep}

\begin{figure}[ht]
    \centering
    \includegraphics[width=.9\linewidth]{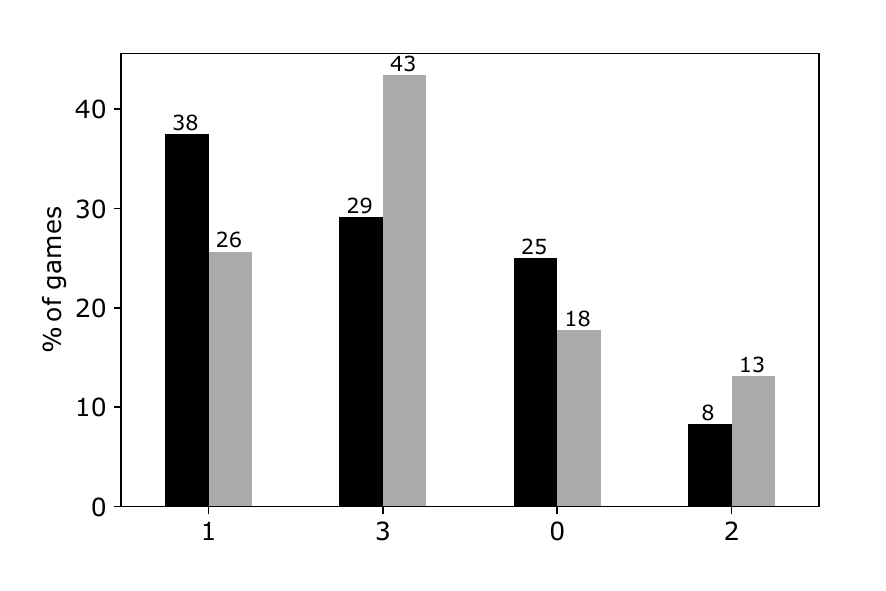}
    \caption{Game \textbf{level of independence} of the high-rated games (dark bars) in comparison with the low-rated ones (gray bars).}
    \label{fig:independence}
\end{figure}

\autoref{fig:independence} shows the levels of independence of the games: \textit{low levels} of independence (levels 0 and 1) occur slightly more frequently in high-rated games; \textit{high-levels} of independence (levels 2 and 3) occur slightly more frequently in low-rated games.

\subsection{Game Genres}
\label{sub:comp-genre}

\begin{figure}[ht]
\centering
\includegraphics[width=1\linewidth]{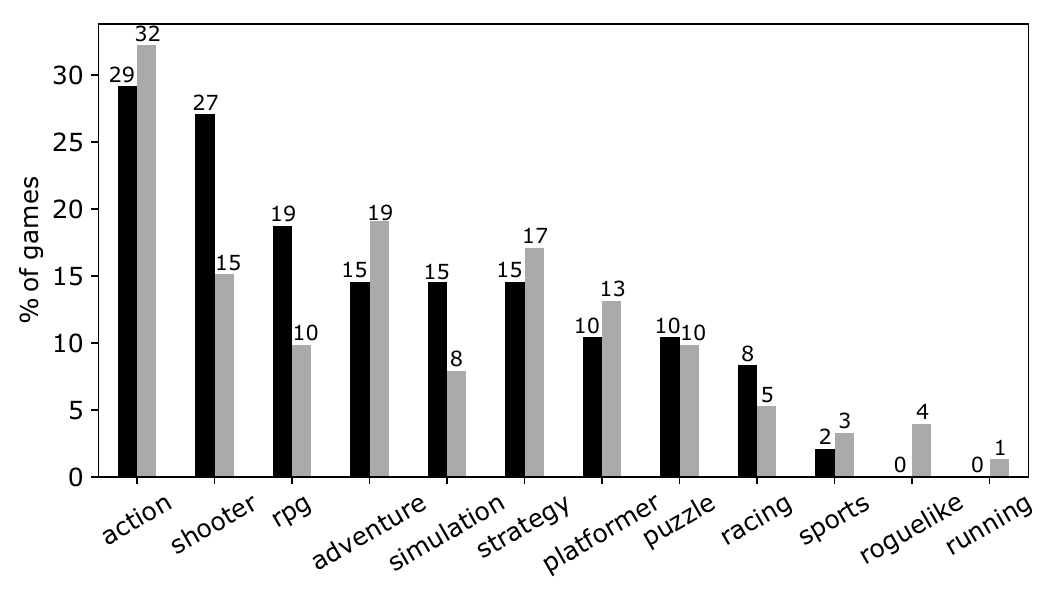}
\caption{Game \textbf{genre} of the high-rated games (dark bars) in comparison with the low-rated ones (gray bars). The percentage is related to each population, not the total (200). Some games have more than one genre.}
\label{fig:genre}
\end{figure}

\autoref{fig:genre} shows the comparison of the games, i.e., high-rated and low-rated, according to their genres. \textit{Shooter}, \textit{RPG}, or \textit{simulation} games are most prone to success, while \textit{action}, \textit{adventure}, \textit{strategy}, and \textit{platform} games are the opposite.

\subsection{Game Modes} 
\label{sub:comp-mode}

\autoref{fig:mode} shows the game modes: \textit{single-player} games are the most common, followed by \textit{multiplayer}. There are very few occurrences of \textit{online-only} games in the dataset. \textit{Multiplayer} games occur slightly more frequently among high-rated than low-rated games. 

\subsection{Game Platforms} 
\autoref{fig:platform} shows that the \textit{PC} is the main platform for games in our dataset. On \textit{PC}, the high-rated games are slightly more common; opposite to what happens in \textit{console} and \textit{mobile} platforms. 

\subsection{Graphical Perspectives} 
\begin{figure*}[ht]
\centering
\begin{subfigure}{.32\linewidth}
    \includegraphics[width=1\linewidth]{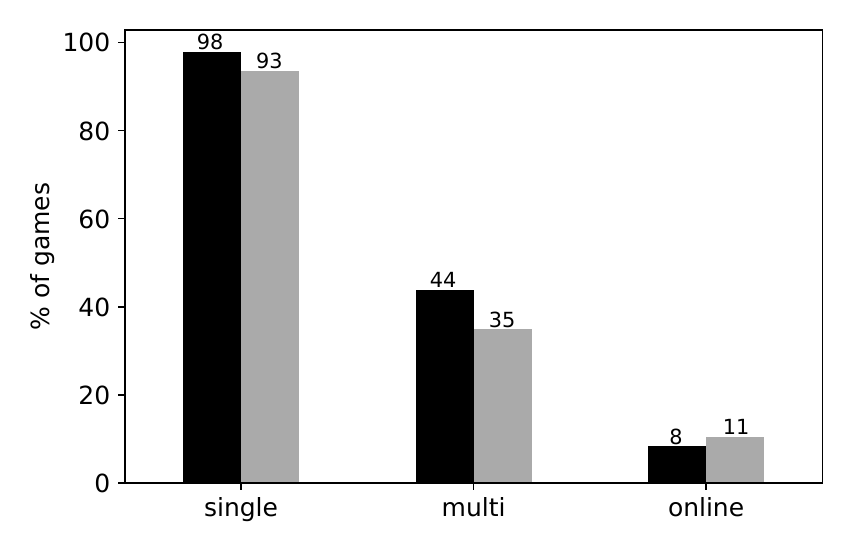}
    \caption{Game \textbf{modes}}
    \label{fig:mode}
\end{subfigure}
\begin{subfigure}{.32\linewidth}
    \includegraphics[width=1\linewidth]{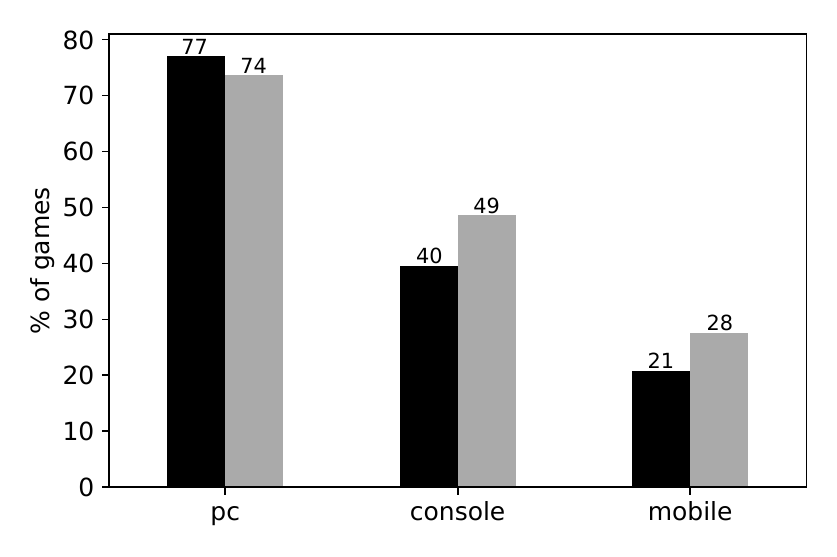}
    \caption{Game \textbf{platforms}.}
    \label{fig:platform}
\end{subfigure}
\begin{subfigure}{.32\linewidth}
    \includegraphics[width=1\linewidth]{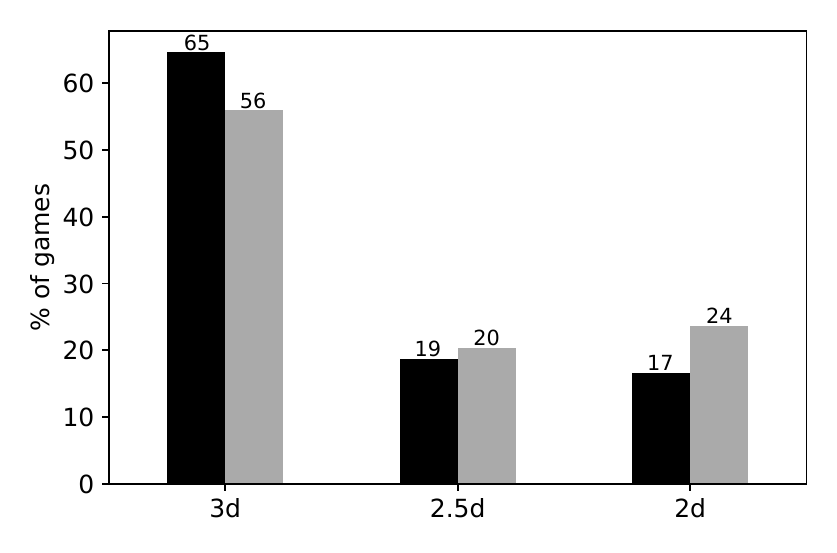}
    \caption{Game graphical \textbf{perspectives}}
    \label{fig:perspective}
\end{subfigure}
\begin{subfigure}{.48\linewidth}
\includegraphics[width=1\linewidth]{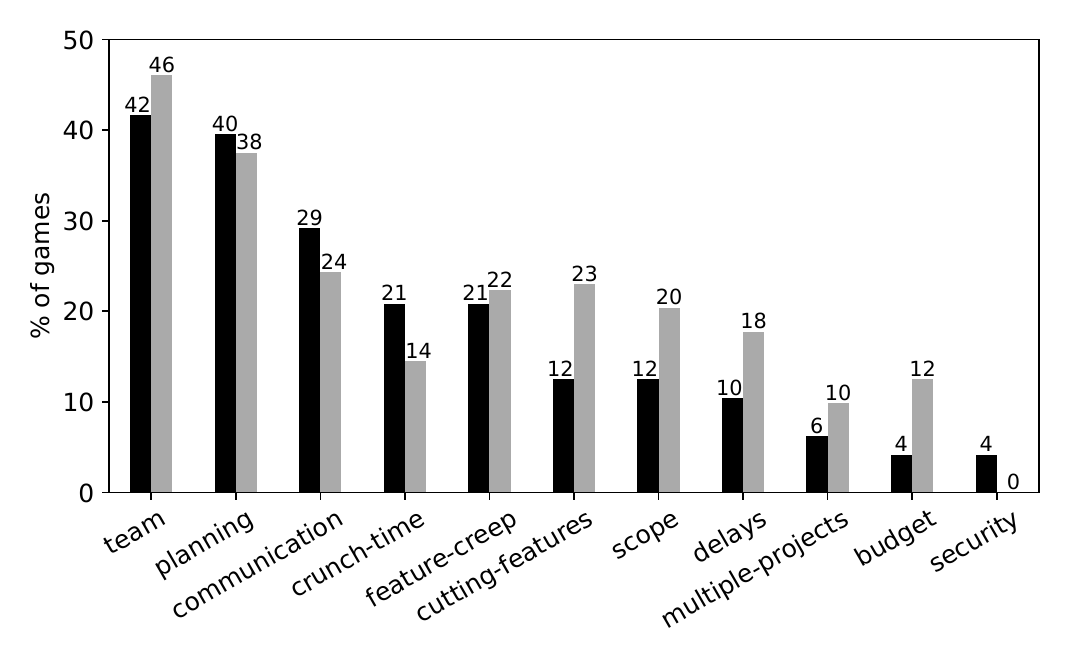}
\caption{\textbf{Management} problems}
\label{fig:prob-mgm}
\end{subfigure}
\begin{subfigure}{.48\linewidth}
\includegraphics[width=1\linewidth]{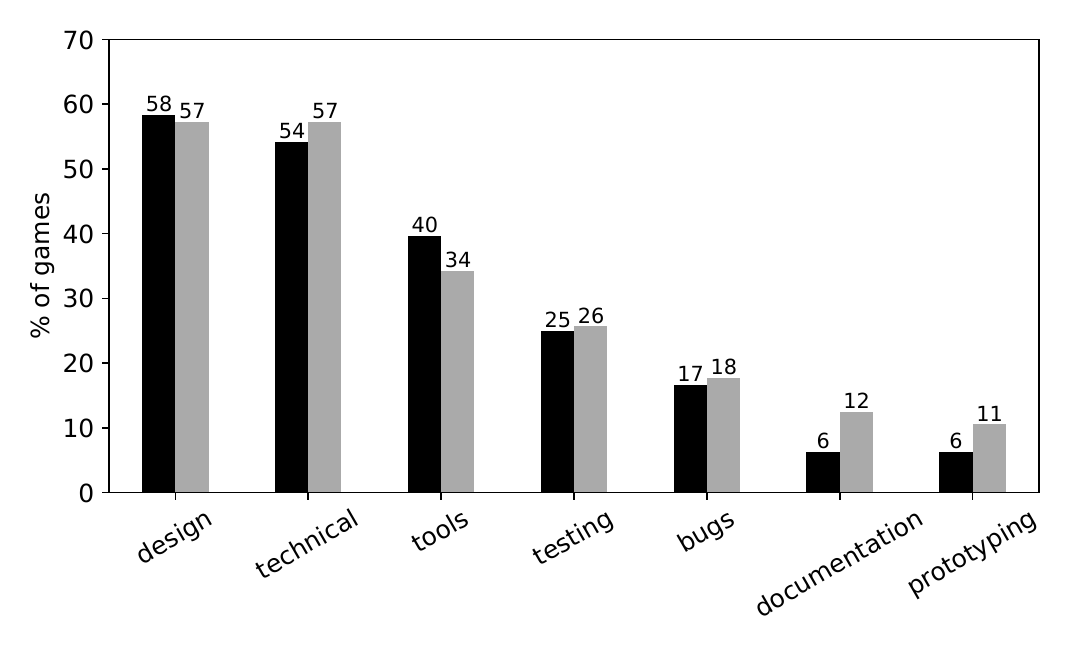}
\caption{\textbf{Production} problems}
\label{fig:prob-prod}
\end{subfigure}
\caption{Factors of high-rated games (dark bars) in comparison with the low-rated ones (in gray bars).}
\label{fig:a}
\end{figure*}

\autoref{fig:perspective} shows the graphical \textbf{perspective} of the games. \textit{3D} games are slightly more common high-rated than in low-rated games. 

\subsection{Problem Types}
\label{sub:comp-prob}

\autoref{fig:prob-mgm} and \autoref{fig:prob-prod} show \textbf{management} and production problems, respectively. The most frequent problems, regardless of score, are \textit{team}, \textit{planning}, and \textit{communication}; \textit{design}, \textit{technical}, and \textit{tools}, respectively. The proportion of problems regarding \textit{cutting-features}, \textit{scope}, \textit{delays} and \textit{budget} is slightly larger for low-rated games. The proportion of \textit{crunch-time} is slightly larger for high-rated games.\linebreak

\bigskip
\begin{summarybox}
Results' summary
\tcblower
Most of the successful games in the dataset are 3D shooters, RPGs or simulations, available for PC in both single and multiplayer mode. These projects usually have teams ranging from 10 to 20 developers. Also, the problems in high-rated and low-rated games are similar. Finally, the freedom developers have does not correlate to better games.
\end{summarybox}

\section{What Went Right}
\label{sec:discussion}

In this section, we explore postmortem quotes to discuss and compare what went right among high-rated projects. \autoref{tab:problem_freq_comp} shows the main factors that ``went right'' and ``went wrong'', the latter being problems. Results show that both high and low rated projects faced similar issues. This corroborates previous findings \cite{politowski20}: \textbf{\textit{team}, \textit{technical}, and \textit{game-design} aspects should be the main focus of the game developers.}

\begin{table}[ht]
\centering
\caption{
The main factors from the WWR sections among the high-rate games. In the right, the problems from the WWW sections \cite{politowski20}, considering all the 200 projects.}
\label{tab:problem_freq_comp}
\begin{tabular}{@{}lrr@{}}
\toprule
\textbf{Factors} & \textbf{WWR (high-rated)} & \textbf{WWW (all projects)} \\ \midrule
Team             & \#1 (60\%)                    & \#3 (08\%)                       \\
Technical        & \#2 (54\%)                    & \#1 (11\%)                      \\
Game Design           & \#3 (42\%)                    & \#2 (11\%)                      \\
Planning         & \#4 (21\%)                    & \#5 (07\%)                       \\
Marketing        & \#5 (19\%)                    & \#6 (06\%)                       \\
Prototyping      & \#6 (13\%)                    & \#18 (02\%)                      \\
Testing          & \#7 (04\%)                     & \#8 (05\%)                       \\ \bottomrule
\end{tabular}
\end{table}

\subsection*{WWR\#1. Team factors}
\label{wwr-team}

Developers mention that \textit{teamwork}, \textit{experienced peers}, and \textit{communication} influence the workflow positively. Others like \textit{familiarity with technology} and \textit{project's vision} are also important for the team's success.

\begin{quoting}[endtext='' -- Command and Conquer: Tiberian Sun by Rade Stojsavljevic]
Several members of the programming team had worked together on previous Westwood RTS products and were accustomed to each other's coding styles.
\end{quoting}

\subsection*{WWR\#2. Technical factors}
\label{wwr-technical}

Game studios often decide to use technologies with which they are familiar, especially when developing sequels to successful games. For example, both the original Baldur's Gate and its sequel were developed with BioWare's Infinity Engine, which was then re-used by publisher Black Isle Studios in several other titles. Also, the studios Stardock (Galactic Civilizations 2) and Outrage (Descent 3) chose not only to keep using proprietary engines from previous games but also to upgrade them:

\begin{quoting}[endtext='' -- Descent 3 by Craig Derrick and Jason Leighton]
Much of the joy of working on a sequel comes from the fact that you can improve on an already established title, and in many cases, add features that were previously impossible to do. (...) One of our biggest goals in developing Descent 3 was to bring the game engine up to date. This included graphics, AI, sound, and multiplayer. I think we hit the mark.
\end{quoting}

\subsection*{WWR\#3. Game-design factors}
\label{wwr-design}

Game design is a broad issue. However, many of the high-rated games cite the search for a ``vision'' or ``blueprint'' as the main element to be addressed and continuously maintained by designers and managers throughout the project. For example, Splinter Cell producer Wu Dong Hao states that his team "regularly followed up on our schedule and made adjustments to [the planning] as necessary to ensure we always had an up-to-date ``vision'' of the project". Alyssa Finley, project lead for Bioshock, states that the project was only successful due to her team's "ability to identify and react when the game was not shaping up to become what it needed to be".

\begin{quoting}[endtext='' -- Startopia by Wayne Imlach]
A past history of successful games did much to strengthen the belief that Startopia's design was an accurate blueprint for a fun and enjoyable game. This highlights the importance of a strong credible design blueprint at the early stage of game development.
\end{quoting}

\subsection*{WWR\#4. Planning factors}
\label{wwr-planning}

The high-rated games described their scheduling as ``flexible'' and ``iterative'', praising their ability to quickly adapt and correct mistakes. Also, having enough time to finish the game as intended is important. For example, Diablo II's PM reports that the team made sure the game was ``as good as it can be before we ship it'' and Prune's PM author says that ``Having the luxury of time allowed me to eventually find the soul of the game''.

\begin{quoting}[endtext='' -- Uncharted: Drake's Fortune by Richard Lemarchand and Neil Druckmann]
Perhaps Naughty Dog's most important achievement is making large-scale games and shipping them on time, with at most a small amount of slip. (...) we prefer a much more flexible, macro-level scheduling scheme, with milestone accomplishments to be achieved by certain dates.
\end{quoting}

\subsection*{WWR\#5. Marketing factors}
\label{wwr-marketing}

Regarding marketing, the main factors of success among the high-rated games are \textit{advertising early}, creating a \textit{strong online presence}, and focusing on understanding the characteristics of the \textit{target public}. Also, getting help from experienced business partners and specialists is an effective way to create a solid distribution, pricing, and sales strategies.

\begin{quoting}[endtext='' -- Galactic Civilizations 2: Dread Lords by Brad Wardell]
One of the best decisions we made was teaming up with Take 2. (...) For Galactic Civilizations II, they were the distributor, not the publisher. Their job was to take our game and put it into stores. (...) We also hired Brian Clair, who had been running Avault.com for many years to be in charge of our publishing efforts. Combining him with Take 2 resulted in having a first week sell-in to retail that was 3 times what the original had.
\end{quoting}

\subsection*{WWR\#6. Prototyping factors} \label{wwr-prototyping}

According to the high-rated games, prototyping early and quickly helped in reaching success. Finding a \textit{solid vision} and \textit{shaping it through iteration} is also important. The postmortems for Guacamelee, Deus Ex, and Prune suggest keeping the team lean and free to experiment during the early iterations.

\begin{quoting}[endtext='' -- Guacamelee by Chris Harvey]
Although questions remained, the game was well defined. This was accomplished using a small group and rapid prototyping, with the freedom for people to try different things.  
\end{quoting}

\subsection*{WWR\#7. Testing factors}
\label{wwr-testing}

QA and testing teams are important in either small and large games. Testing might also involves end-users, which can help creating a community, as the postmortems from Myth: The Fallen Lords, Dark Age of Camelot, Bionic Commando Rearmed, and Armadillo Run recommend.

\begin{quoting}[endtext='' -- Myth: The Fallen Lords by Jason Regier]
There's no big quality assurance department here at Bungie; the public did our testing for us, and we listened to them as seriously as if they were coworkers on the project.
\end{quoting}

\bigskip
\begin{summarybox}
Discussions' summary
\tcblower
The main suggestions from high-rated projects are: follow a well-defined design \textit{vision}, take the \textit{time} to develop the game, and control the game \textit{scope}. The development \textit{team}, technical skills (experienced and well-balanced team) as well as soft skills (close-knit team) are equally important.
\end{summarybox}

\section{Threats to Validity}
\label{sec:threats}

\paragraph*{Scores} 
The game scores were collected from multiple sources that used different scales. Normalizing these values to a common scale is not as precise as using only one source. Also, differently from Metascore, Steam, Google Play and App Store ratings are based on user and not critic reviews, therefore being vulnerable to "review bombing". However, there are no known occurrences of "bombings" against any of the games in our dataset.

\paragraph*{Team sizes} 
Not all sources of information offered a precise description of their teams, in terms of numbers and composition. While some focused on describing mostly technical personnel or key people to the project (e.g., leads and managers), others mentioned people from a broad range of positions and departments.

\paragraph*{Levels of independence}
Many postmortems refrain from discussing all phases of the project in detail and frequently offer vague information related to specific management and financial decisions. Therefore, it is possible that the occurrence of financial and creative independence could be wrongly inferred.

\paragraph*{Game genres}
Given the fact that any game could be classified into one or more genres or sub-genres, our genre definitions cannot be taken as absolute.

\paragraph*{Graphical perspectives} 
Considering that there is no standard definition for graphical perspective in the context of games, our classification cannot be taken as absolute. Some games could even fit more than one definition (e.g: switching between 2d and 3d views).

\section{Related Work}
\label{sec:related}

Many works have analyzed the factors of success for games through distinct approaches. Empirical studies, for example, evaluate marketing, monetization, brand and customer service strategies \cite{Aleem2016} \cite{cabras17}, as well as company organizational structure, location \cite{tozour14} and publishing plan \cite{thomes2015} to infer the success of its projects. 

Other studies investigate success from the point of view of the public and critics, analyzing scores provided by specialized websites such as Metacritic or Gamespot \cite{bond09}, as well as Steam text reviews written by players \cite{lin18}. The latter is analyzed in conjunction with the reviewer's playing hours as a way to infer the level of player engagement.

By studying failure/success cases through postmortem documents \cite{petrillo10} or analyzing a specific area of game development such as QA \cite{yarwood20}, some works also aim to describe good practices or even propose completely new project frameworks, built to avoid mistakes that occur in well-established approaches \cite{ahmad2017}.

Some works used the ``what went right'' sections in the postmortems to create a set of good practices. First, \citet{petrillo10} made a parallel with the agile practices. Later on, \citet{Washburn2016} also created a list of the most common practices.

\section{Conclusion}
\label{sec:conclusion}
In this work, we investigated which aspects better describe the high-rated games. We collected data from 200 video game projects and analysed a set of factors: \textit{team size}, \textit{levels of independence}, \textit{game genre}, \textit{game platform}, \textit{graphical-perspective}, \textit{game mode} and \textit{development problems}. We also collected numeric scores for each game using ``Metacritic'' and other review aggregators. We used this data to investigate which of these factors affect the success of the games. 

We asked the research question: \textbf{Which factors better describe high-rated games?} We answer that while none of the factors have a strong relationship with games success, our results show that larger teams are often related to higher-rated games. Higher-rated games also present more occurrences of crunch time but less problems with scope, delays, budget, and cutting features. Additionally, the level of freedom given to game developers does not correlate to scores.

A further analysis on the games confirmed previous study \cite{politowski20} in that \textit{team}, \textit{technical}, and \textit{game-design} factors should be the main focus of the game developers. These findings may serve as guidelines for video game project managers regarding important points to be monitored in any project, from planning to execution. 

\paragraph{Limitations} Over the last decades, the video game market has been growing and there has been a constant influx of new gaming platforms and titles every year. Given this broad scenario, our dataset may be too small to represent the frequency of game development issues.

Also, our choice of measuring success through scores can be argued to be an oversimplification. Review aggregators may serve as an indicator of how much the critics or the audience enjoyed a given title, but this may not be directly linked to its financial success, or the performance of a studio's team and project management practices.

\paragraph{Future Work} In future works, we aim to expand our database, adding to our analysis a larger number of postmortems. Additionally, we will study the aforementioned factors in-depth, to understand how to apply better practices to game development. For example, how to balance the skill expertise on a given video game development team? Does it change depending on the game genre? How to keep a design vision given the iterative nature of the game development? Many challenges will require specific studies to be addressed properly.

\bibliographystyle{ACM-Reference-Format}
\bibliography{main.bib}

\end{document}